\documentclass[10pt,onecolumn,twoside,notitlepage]{revtex4-1}

\usepackage[utf8]{inputenc}

\usepackage{amsmath}
\usepackage{subfigure}
\usepackage{graphicx}
\usepackage[pagebackref=true]{hyperref}

\begin{document}





\title{Coincidence detection of spatially correlated photon pairs with a monolithic time-resolving detector array}

\author{Manuel Unternährer$^{1,*}$, Bänz Bessire$^1$, Leonardo Gasparini$^2$, David Stoppa$^2$, and André Stefanov$^1$}
\affiliation{$^1$ Institute of Applied Physics, University of Bern, 3012 Bern, Switzerland\\
$^2$ Fondazione Bruno Kessler FBK, 38122 Trento, Italy\\
$^*$ Corresponding author: manuel.unternaehrer@iap.unibe.ch}


\begin{abstract}
We demonstrate coincidence measurements of spatially entangled photons by means of a multi-pixel based detection array. 
The sensor, originally developed for positron emission tomography applications, is a fully digital 8$\times$16 silicon photomultiplier array allowing not only photon counting but also per-pixel time stamping of the arrived photons with an effective resolution of 265\,ps. Together with a frame rate of 500\,kfps, this property exceeds the capabilities of conventional charge-coupled device cameras which have become of growing interest for the detection of transversely correlated photon pairs. The sensor is used to measure a second-order correlation function for various non-collinear configurations of entangled photons generated by spontaneous parametric down-conversion. The experimental results are compared to theory.\\
\\
\doi{10.1364/OE.24.028829}
\\
\\
© 2016  Optical Society of America. One print or electronic copy may be made for personal use only. Systematic reproduction and distribution, duplication of any material in this paper for a fee or for commercial purposes, or modifications of the content of this paper are prohibited.
\end{abstract}

\maketitle



\section{Introduction}
The non-linear interaction of spontaneous parametric down-conversion (SPDC) has become a pervasive process to obtain spatially entangled two-photon states used in experiments addressing fundamental properties of quantum mechanics as well as practical applications \cite{walborn10}. Double-slit induced interference patterns using transversely entangled photons were studied in \cite{ribeiro94,walborn02,brida03,peeters09}. Continuous variable entanglement in the spatial degrees of freedom of the photon pairs was demonstrated in \cite{reid09,moreau12,edgar12,moreau14} and transverse entanglement in ghost imaging has been shown to create a spatially resolved image of an object carried by a photon which did not interact with the object itself \cite{lugiato02,shapiro12,aspden13,genovese16}. Finally, due to their inherent high-dimensionality, the transverse degrees of photons are also a primary resource to perform quantum information tasks. Entangled $d$-dimensional qudit states were implemented using transverse spatial correlations \cite{neves05}, in a discrete set of orbital angular momentum modes in \cite{mair01,vaziri02,nagali10,dada11,agnew11} or in the intensity profile of Hermite- or Laguerre-Gauss modes \cite{langford04,salakhutdinov12,krenn14}. Transverse correlation based protocols for quantum key distribution and teleportation were realized in \cite{almeida72} and proposed in \cite{walborn07}. 

By doing so, the detection of transverse photon correlations has been subjected to change in the last few years. Past experiments resolved these correlations by scanning apertures in front of single photon detectors while measuring a position dependent correlation function. Such detection schemes were deployed from the very first experiment investigating the spatio-temporal properties of SPDC photons \cite{burnham70}, to the concept of ghost imaging \cite{pittman95} and to the early detection of Einstein-Podolsky-Rosen (EPR) correlations in the transverse position and momentum of the entangled photons \cite{howell04}. 

However, to overcome the time consuming scanning process, parallel detection by multi-pixel arrays has become the preferred method to record coincidences of spatially entangled photons. Thereby, as conventional charge-coupled device cameras (CCDs) cannot work in the photon-counting regime, mainly due to readout-noise,
and are not capable of sub-nanosecond time resolution,
optical and electrical amplification schemes are used. In both of these systems, time resolution for coincidence detection is determined by the shutter/gating time or pulsing of the light source. Electrical amplification is implemented by electron-multiplying CCDs (EMCCDs) where an amplifier stage, using avalanche diodes, enhances the collected photoelectrons before the output amplification and analog-to-digital conversion. The accompanying drawbacks are the costly cooling needed and the large gating time window of the order of microseconds \cite{zhang09}. Because of the latter, low SPDC fluxes were used to investigate spatial correlations \cite{zhang09}, photon statistics \cite{blanchet08} and EPR-type entanglement \cite{edgar12, moreau14} by means of EMCCDs. Further, an absolute calibration of an EMCCD was performed in \cite{avella16} using spatially entangled photon states.
The optical amplification approach uses an image intensifier, consisting of a photocathode, a multi-channel plate and a phosphor screen, in front of a CCD. 
By reversing the voltage on the photocathode, intensifiers can be gated for sub-nanosecond time windows and therefore are not relying on low fluxes or pulsed sources. Such intensified charge-coupled devices were used to study spatial correlations in SPDC \cite{jost98, oemrawsingh02}, spatial entanglement \cite{dilorenzo09, fickler13} and ghost imaging \cite{aspden14, morris15}.

A different type of sensor array is  used in \cite{just14}. This hybrid detector is a CMOS integrated circuit developed for electron detection in particle physics and is used in combination with a photomultiplier. It exhibits a high spatial resolution given by $256\times 256$ pixels which stores the time of the first detection event in a frame with a resolution of 10\,ns. Coincidence measurements with SPDC light allowed to determine its detection efficiency. Further, a monolithic array of single-photon avalanche diodes (SPADs) based on CMOS technology is used in \cite{boiko09} to spatially resolve second-order intensity correlations in order to measure temporal correlation functions of classical light. Recently, several other SPAD-based CMOS sensors have been reported for time-resolved single-photon applications such as: fluorescence lifetime imaging \cite{burri14}, time-resolved fluorescence spectroscopy \cite{krstajic15} or 3D time-of-flight imaging \cite{lussana13,jahromi16}. These SPAD-based sensors are good potential candidates for coincidence detection of spatially correlated photon pairs, however, they suffer from very low fill factors \cite{burri14,lussana13}, time-coincidence detection capability longer than 0.6\,ns \cite{burri14,krstajic15,lussana13}, acquisition frame rates below 160\,kHz \cite{burri14,krstajic15,lussana13,jahromi16}, limited spatial resolution of the 2D pixel array arrangement \cite{burri14,krstajic15}. The state-of-the-art in terms of spatial resolution for a SPAD-based sensor is represented by \cite{dutton16} that reports a QVGA 8-$\mu$m pixel pitch with 26.8\% fill factor, however its time-resolving performance is in the order of nanoseconds and not fully reported yet while the sensor frame rate is of 16\,kfps. 

In this work, we demonstrate coincidence detection of spatially correlated photon pairs by means of the \mbox{SPADnet-I} sensor, a 8$\times$16 pixel single photon detector based on CMOS-technology \cite{braga14}. 
\mbox{SPADnet-I} converts the SPAD signal from the analog to the digital domain at pixel level, thus avoiding spurious correlations due to inductive wire coupling. The \mbox{SPADnet-I} pixels are individually equipped with time-to-digital converters (TDCs). This allows for per-pixel timestamping of the detected photons with 265\,ps resolution. Furthermore, the detector frame rate of 500\,kfps outperforms conventional CCD based camera systems by at least three orders of magnitude and surpasses the frame rates of the aforementioned SPAD-based CMOS sensors in \cite{burri14,krstajic15,lussana13,jahromi16,dutton16}. The here presented fill factor of 42.6\% moreover exceeds the fill factors presented in \cite{boiko09,burri14,lussana13,dutton16}. By means of non-classical light states generated by continuous wave SPDC we demonstrate the ability of the here presented sensor to measure a second-order correlation function for various non-collinear propagation modes of the photons. Thereby, we compare the experimental results to theoretical predictions.

\section{SPADnet-I sensor}

\mbox{SPADnet-I} is a fully digital silicon photonic device based on SPAD arrays implemented in a 130\,nm CMOS technology. It consists of an 8$\times$16 array of pixels of 610.5$\times$571.2\,$\mu$m$^2$ area, for a total size of 9.85$\times$5.45\,mm$^2$. Each pixel contains 720 SPADs of circular shape with a diameter of 16.87$\mu$m, the electronics required to count photons and two 12-bit TDCs each having a nominal time resolution of 65\,ps. SPAD detection time jitter as well as electronic jitter reduces the effective resolution to 265\,ps. 
The SPADs can be individually enabled and disabled due to a dedicated programmable 1-bit memory cell. This is typically done for those exhibiting a high dark count rate (DCR), i.e.~a high rate of avalanche events induced by thermal generation or tunnelling rather than photon detection. A photon detection efficiency (PDE) of 19\% is achieved at the design wavelength of 450\,nm, whereas at the wavelength of 810\,nm used in this work  a PDE of 1.1\% is reported \cite{braga14}.

\begin{figure}[t]\
 \centering
\includegraphics[width=.8\textwidth, trim=1cm 0 0 0]{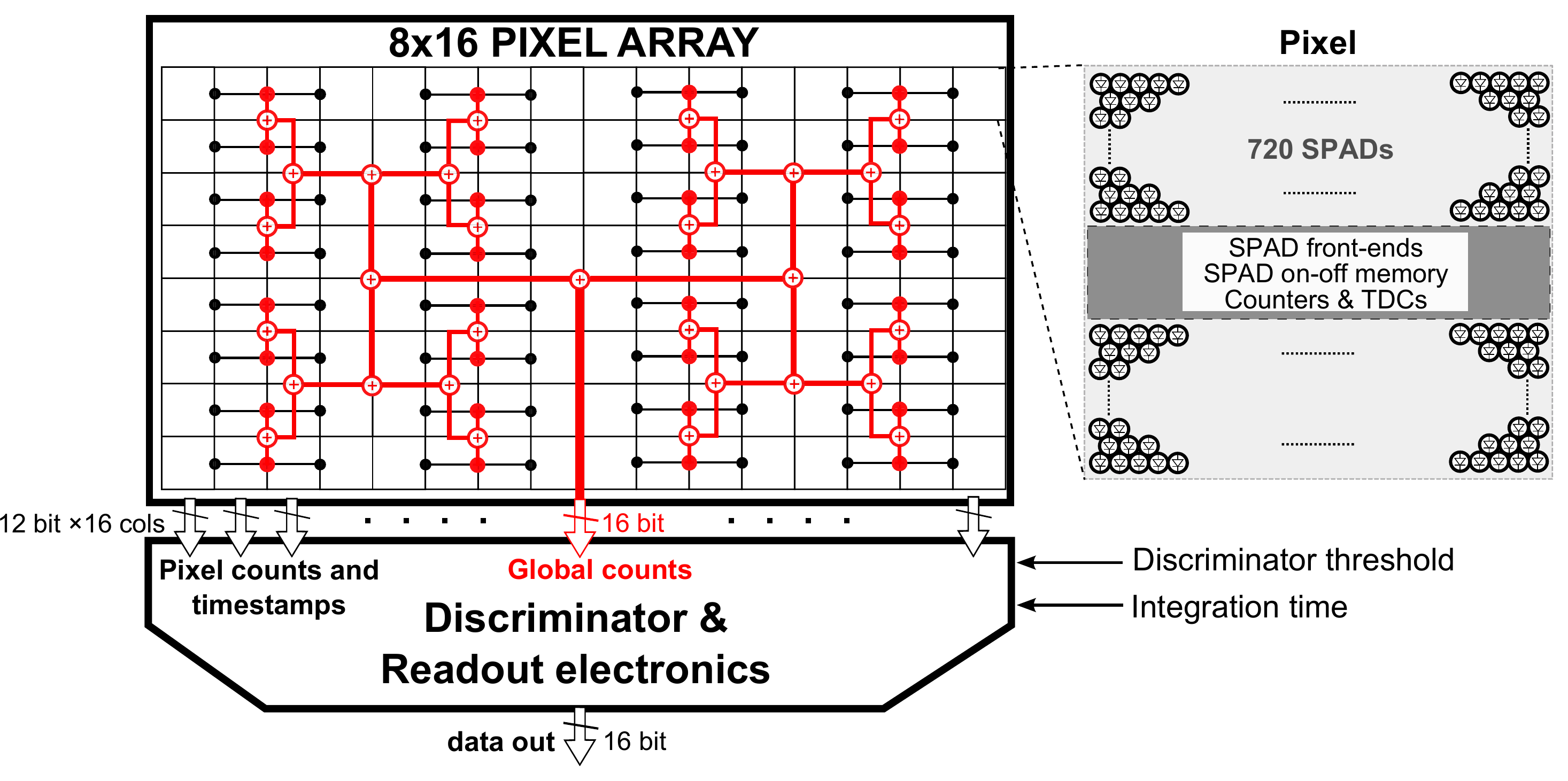}		
\caption{Architecture of the \mbox{SPADnet-I} sensor. It includes an 8$\times$16 array of pixels, each including 720 SPADs, photon counters and TDCs. A tree of adders is distributed across the array to calculate the number of triggering SPADs at 100\,MHz. Additional logic units are present at the periphery of the array for event discrimination and data readout. Operations are synchronous with a global clock.}
\label{fig:SPADnet_arch}	
\end{figure}

The sensor is synchronous with a global clock signal that can be operated at up to 100\,MHz. For every clock bin, each pixel generates a photon count (number of the SPADs triggered in the current bin) and one  photon timestamp of the first photon detected in the bin.
At the same rate, a distributed network of adders computes the number of photons detected globally. Fig.~\ref{fig:SPADnet_arch} shows the architecture of the chip.

\begin{figure}[t]\
  \centering
  \includegraphics[scale=.8, trim=0.3cm 0 0 0]{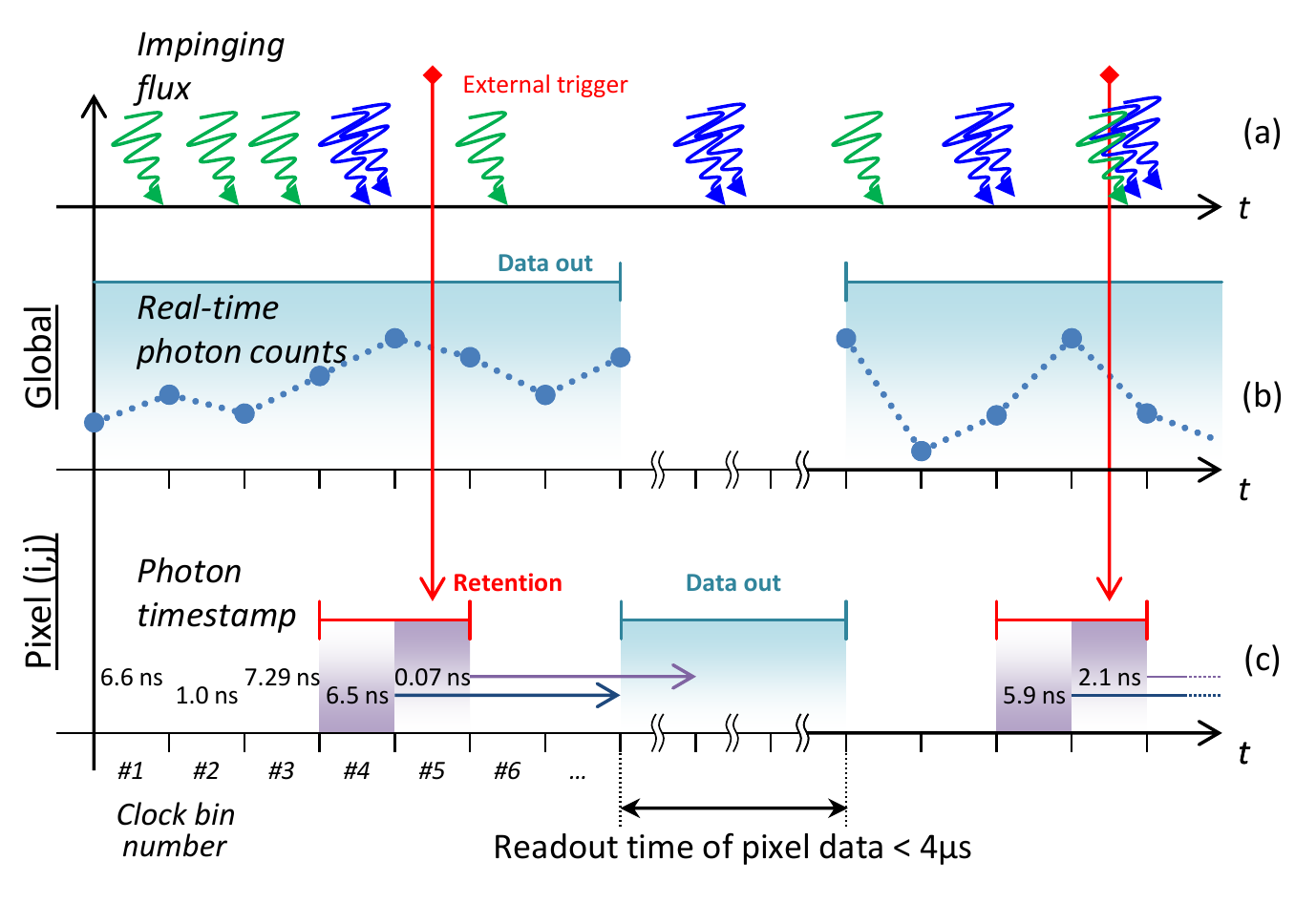}
\caption{Timing diagram of the \mbox{SPADnet-I} sensor operation  with example data, adapted to quantum optics experiments, looking for coincident photons (blue arrows in \textit{a}). During standard operation (clock bins \#1-\#3) each pixel generates photon counts (not shown) and timestamps (\textit{c}) at the clock rate, while the sensor streams out the number of photons globally detected (\textit{b}, light blue area). When the external trigger is provided (vertical red arrows) each pixel retains the photon time-stamps (\textit{c}, purple areas), the stream of global counts is interrupted and data are read out of the chip (\textit{c}, light blue area). Then the process starts over again. The sensor is read out at the maximum frame rate limited by readout time.}
    \label{fig:SPADnet_QO}
\end{figure}

\mbox{SPADnet-I} has been specifically designed for gamma ray monitoring in positron emission tomography (PET) applications and includes additional logic for this purpose~\cite{braga14}. In a PET system, sensors are coupled to crystal scintillators that convert gamma photons into bursts of visible photons. The capability of recording the photon arrival time with a relatively high spatial resolution (with respect to analog silicon photo-multipliers), in conjunction with a high fill factor (with respect to other CMOS SPAD arrays with per-pixel time-stamping capabilities) of 42.6\% makes \mbox{SPADnet-I} suitable for quantum optics applications. 
In this context, the sensor is read out using an external trigger at a fixed rate of up to 250\,kHz limited by readout time, see Fig.~\ref{fig:SPADnet_QO}.
Every data acquisition provides for all pixels the timestamps of two consecutive clock bins. Therefore, 8$\times$16 maps of photon timestamps (hereinafter referred to as \textit{frames}) are generated at up to 500\,kfps.
A timestamp is an integer TDC code which is thereby measured in TDC units of 65\,ps. The measurement or exposure time of one frame is given by the period of the global clock signal.

Crosstalk events are spurious, simultaneous detection events between pixels. Since the digital signal handling at pixel level prevents electrical crosstalk, mainly photonic crosstalk is expected: light emitted in a SPAD avalanche event leads to secondary detection events in neighbouring pixels. The temporal and spatial correlation of these events will be present in the following measurements.

\section{Experiment}\label{Sec:SPDCLight}
\begin{figure}[tbp]
\includegraphics[width=1\linewidth]{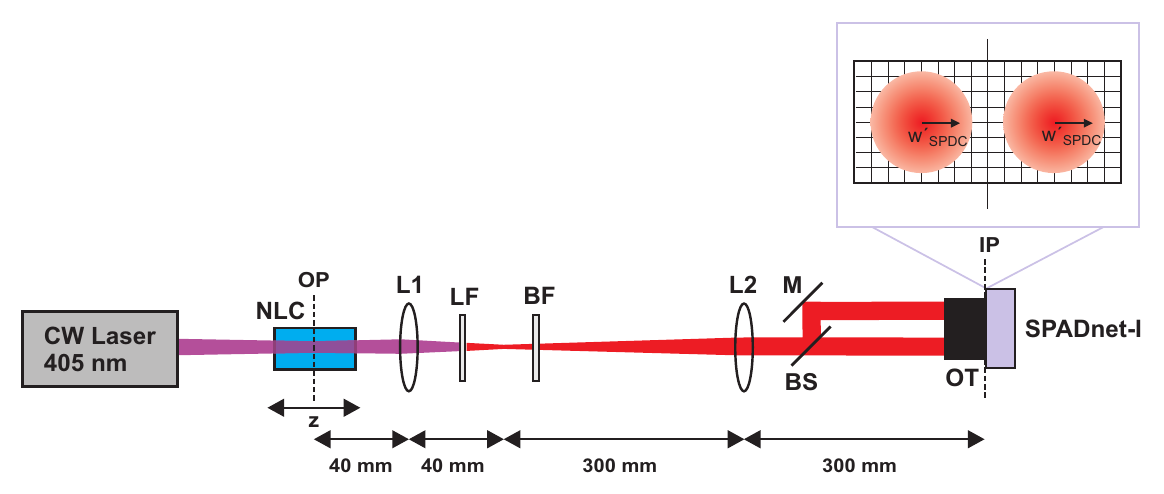}
\caption{Experimental setup. CW continuous wave pump laser at 405 nm, NLC non-linear crystal movable in $z$-direction, L1 lens ($f_1=40$\,mm), LF longpass filter to reject the residual of the pump, BF bandpass filter (810\,nm, 10\,nm FWHM), L2 lens ($f_2=300$\,mm), BS 50:50 plate beam splitter, M mirror (the distance between BS and M is 2\,cm), OT optically opaque tube to reduce the effect of stray light, SPADnet-I sensor. The telescope (L1,L2) provides a magnification of $m=8$ from the object plane (OP) to the imaging plane (IP). The inset shows the arrangement of the two beams on the sensor surface. The magnified beamwaist is $w'_{SPDC}\approx 2$ mm and covers about 3 pixels in radius.}\label{fig:expsetup}
\end{figure}

The experimental setup is depicted in Fig.~\ref{fig:expsetup}. Spatially entangled photon pairs are created by degenerated type-0 SPDC in a 12\,mm long KTiOPO$_4$ (PPKTP) non-linear crystal (NLC) pumped by a quasi-monochromatic laser operating at $\lambda_{p,c}=405$\,nm with a power of 33\,mW. The pump ($p$) beam is focused into the middle of the NLC with a beam waist of $w_p=0.25$\,mm. The residual pump beam is afterwards filtered out by a longpass filter and a subsequent bandpass filter transmits photons at 810\,nm with central frequency $\omega_c=\omega_{p,c}/2$. The corresponding biphoton state can be derived by perturbation theory under the assumption of a classical plane-wave pump field and a fixed central frequency $\omega_c$. The first-order order correction to the vacuum state then reads 
\begin{equation}\label{eq:state_vec}
\vert \Psi \rangle = \int d^2q \; \Lambda(\mathbf{q},-\mathbf{q}) \;
\vert 1_{\mathbf{q}} \rangle_s\ \vert 1_{-\mathbf{q}} \rangle_i,
\end{equation}
where $\mathbf{q}=(q_{x},q_{y})$ denotes the transverse momentum of the signal ($s$) and idler ($i$) photon \cite{walborn10}. The transverse joint momentum amplitude $\Lambda(\mathbf{q},-\mathbf{q})$ governs the phase matching condition of the SPDC process and is, for the approximations used to derive Eq.~(\ref{eq:state_vec}), explicitly given by
\begin{equation}\label{eq:sinc}
\Lambda(\mathbf{q},-\mathbf{q}) \propto \mathrm{sinc}\left\lbrace{\frac 1 2 \left[\Delta k_{z}(\mathbf{q},-\mathbf{q},\omega_c,T)+\frac{2\pi}{G}\right] L} \right\rbrace
\end{equation}
with the NLC  length $L$ and poling period $G$.
The phase mismatch $\Delta k_{z}(\mathbf{q},-\mathbf{q},\omega_c,T)=k_s(\mathbf{q},\omega_c,T)+k_i(-\mathbf{q},\omega_c,T)-k_p(0,2\omega_c,T)$ includes the dispersion characteristics of the NLC through its temperature dependent Sellmeier equation \cite{emanueli03}. The minor additional temperature dependence of $\Lambda$  due to thermal length expansion of $G$ and $L$ is negligible in our configuration. In the experiment, the crystal temperature is stabilized to $\pm 0.01^\circ$C and allows to modify Eq.~(\ref{eq:sinc}) for collinear and non-collinear emission. Additionally, the crystal position can be varied in $z$-direction by means of a manually driven linear stage. We experimentally determined a poling period of $G=3.511$\,$\mu$m in a separate measurement with fibre coupled detectors where the dependence of the near-field coincidence rate on the NLC temperature $T$ was measured.

In order to better separate the effective coincidence signal from unwanted crosstalk between adjacent pixels we split the entangled photon beam into two beams which are then arranged next to each other on the detector. Additionally, this allows to detect coincidence events between photons which are spatially separated below the size of a single pixel. To image the entangled photons from the object plane to the SPADnet-I sensor such that the two adjacent beams cover a large area on the sensor without overlapping we choose a telescope system using two lenses with focal lengths $f_1=40$\,mm and $f_2=300$\,mm. The measured magnification factor is $m=8$. The SPDC photon pairs are spatially distributed across the transverse pump profile and thus for their waist $w_{SPDC}$ it holds that $w_{SPDC}\approx w_p$. Given the magnification of the used telescope, this leads to a beam waist of the entangled pairs of $w'_{SPDC}=mw_{SPDC}\approx 2$\,mm at the imaging plane which coincides with the active surface of the sensor. Therefore, the magnified beam waist of the SPDC photons covers about 3 pixels in radius in each half of the sensor (Fig.~\ref{fig:expsetup}). The beam separation itself is performed by means of a 50:50 plate beam splitter and a mirror in front of the imaging plane.
The length difference of 2\,cm of the reflected beam defocuses the image only to a small extent and is unobservable in our measurements due to the limited spatial resolution of the sensor given by the pixel size.
Additionally, the active area of the sensor is shielded with an optically opaque tube to reduce the detection of stray light. The flux of entangled photons impinging on the detector is 2.1\,nW which corresponds to $ 8.6\times 10^9$\,ph/s. Finally, the sensor is connected to a PC via Ethernet and data gathering is performed using LabView.

Coincidence events between distant pixels are described by a second-order correlation function 
\begin{equation}\label{eq:g2}
\begin{split}
G^{(2)}(\pmb{\Delta\varrho},z) & \propto\bigg\vert \int d^2q \; \Lambda(\mathbf{q},-\mathbf{q}) \\
& \times H_{s}(\mathbf{q},z)H_{i}(-\mathbf{q},z)\exp(-i\mathbf{q}\,\pmb{\Delta\varrho}/m) \bigg\vert^2,
\end{split}
\end{equation}
where $\pmb{\Delta\varrho}=\pmb{\varrho}_1-\pmb{\varrho}_2=(\Delta x,\Delta y)$ denotes the distance between the transverse positions $\pmb{\varrho}_1$ and $\pmb{\varrho}_2$. Further, $m$ is the magnification factor of the imaging system. The transfer function $H_{j}$, $j\in\{s,i\}$, describes an additional free space propagation of the signal (idler) photon along a distance $z$ which is equivalent to move the crystal in $-z$ direction (Fig.~\ref{fig:expsetup}). The corresponding transfer function in paraxial approximation is given by
\begin{equation}
H_{j}(\mathbf{q},z)=\exp\left[-ikz +\frac{iz}{2k}\vert\mathbf{q}\vert^2\right],
\end{equation}
where $k=\omega_c/c$. Note that by our specific choice of coordinates, $z=0$\,mm fixes the object plane in the middle of the NLC.

\section{Results}
In the following measurements, the \mbox{SPADnet-I} clock signal is operated at 100\,MHz leading to a measurement time of 10\,ns per frame. The frame readout rate is set to 330\,kfps. This corresponds to measurement duty cycle of 330\,kHz$\times$10\,ns = 0.33\%.
If not stated otherwise, the acquired data consists of 5.4\,M frames corresponding to 1.3\,GB raw binary data which could be acquired in 16~seconds given the mentioned frame rate. At this data rate ($\sim$0.5\,Gbit/s), the limited computational performance of the PC for the real time analysis prolongs the measurement to 45~seconds due to dropped and thereby lost frame data. The effective measurement time of 5.4\,M frames, during which the sensor acquires time-resolved detection events, is $5.4\,\text{M}\times 10\,\text{ns} = 54\,\text{ms}$. To reduce the dark counts, 50\% of the highest DCR SPADs are disabled. Crosstalk between pixels is suppressed by further turning off SPADs in the boundary region between two pixels, leaving a gap of $\sim70\,\mu$m between them. In total, a fraction of 36\% of all SPADs are used in the subsequent measurements.

\subsection{Single photon detection}\label{Sec:SinglePD}
\begin{figure}[tb]
   \subfigure[]{\includegraphics[height=3.5cm]{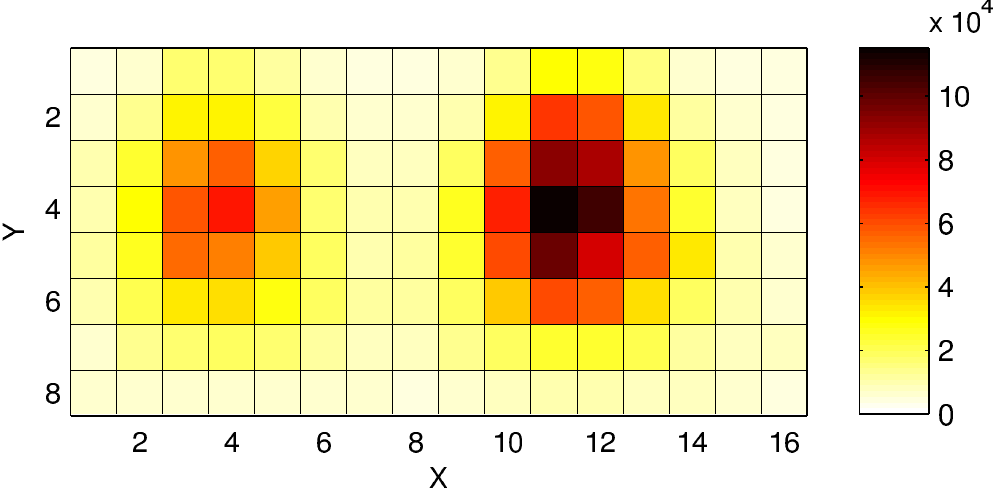}\label{fig:Singles}}\hspace{1cm}
  \subfigure[]{\includegraphics[height=3.5cm]{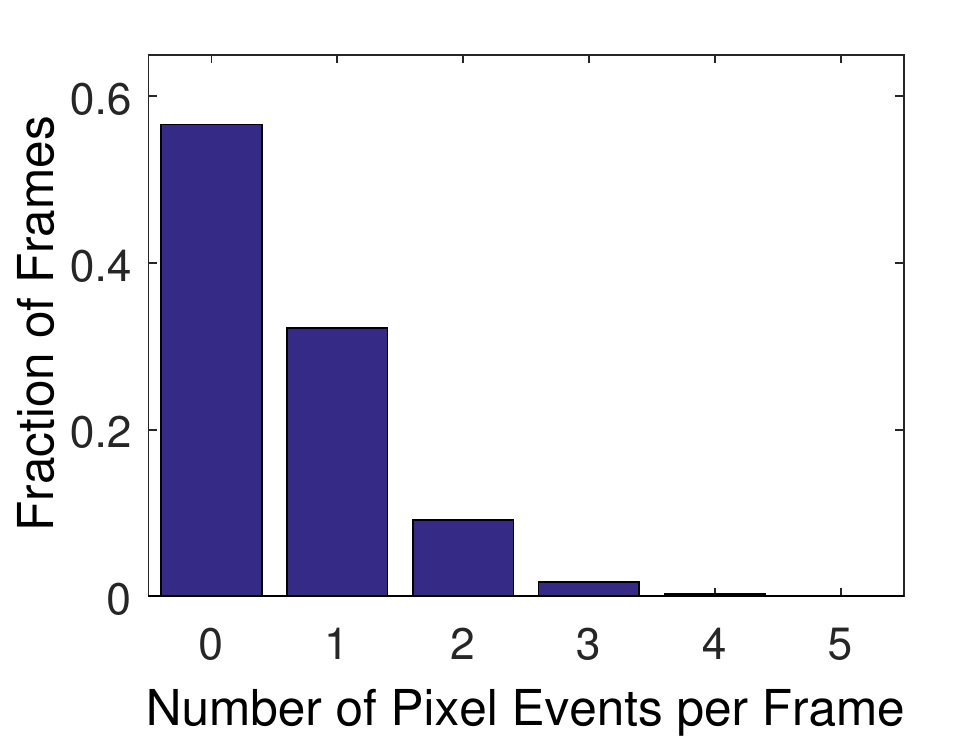}\label{fig:DetPerTDCI}}
    \caption{Single photon counting events. Panel (a) depicts the sensor pixel array with the number of detection events per pixel. A total of 3.07\,M events are registered in 5.4\,M frames. 
    The intensity of the left hand beam is slightly degraded due to the non-perfect 50:50 behaviour of the beam splitter. Panel (b) shows the distribution of the total number of detection events in a frame.}
\end{figure}

Figure \ref{fig:Singles} shows the spatially resolved total number of single photon detection events at maximal SPDC power. The shown maximal photon detection number of the order of $10^5$ per pixel leads to a maximum of 0.02 registered events per pixel and per frame. Taking into account all pixels we measure an average of 0.56 events per frame. Figure \ref{fig:DetPerTDCI} depicts a distribution of the number of detection events $N$ per frame where it can be seen that 55\% of the frames contain no event. By taking into account all intervals with $N\geq 1$ we obtain a total number of 3.07\,M single photon detection events.
Measuring the same number of frames without any incident light, 
a total of 427\,k dark count events are registered. They are homogeneously distributed across all pixels. Per frame, this translates to an average of 0.08 events in total and a maximum of 0.0006 events per pixel.
At the mentioned incident photon flux, the detection number corrected for dark counts, and effective acquisition time, the photon detection efficiency (PDE) is  $0.57\%$, or $1.6\%$ if we correct for the amount of disabled SPADs. This is slightly higher then the PDE of 1.1$\%$ at 810\,nm reported by \cite{braga14}. The DCR of all pixels is 7.9\,Mcps.

\subsection{Coincidence detection}\label{AccRemoval}
\begin{figure}[tp]
  \centering
  \includegraphics[scale=.8, trim=.5cm 0 0 0]{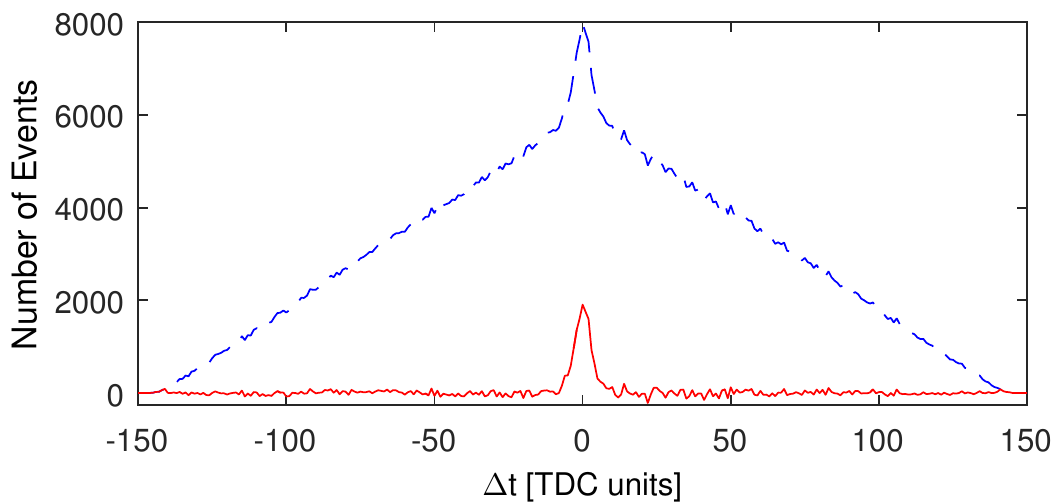}  
    \caption{Histogram of the time difference $\Delta t$ between all events within every frame. Photon pair detections are expected at $\Delta t = 0$\,TDC units. The raw data (dashed line) shows a linear background of accidental events which are also present in a coincidence window around $\Delta t=0$\,TDC units. These accidental events are removed by linear fitting and extrapolation (solid line). 1\,TDC unit $\approx 65\,ps$.}
    \label{fig:TimeDiffHisto}
\end{figure}

All intervals with $N\geq 2$ in Fig.~\ref{fig:DetPerTDCI} are considered to evaluate coincidence events. A histogram of the time differences between all events within every frame is shown in Fig.~\ref{fig:TimeDiffHisto}. A coincident detection of a photon pair is expected to appear at small time differences $\Delta t$ due to a coherence time of the entangled photons of about 500\,fs. The histogram reveals, on top of a linear background of accidentals, a peaked signal with a FWHM of 6\,TDC units which corresponds to $\sim$390\,ps. From pixel-to-pixel, the FWHM variation of the TDC unit of 64.56\,ps is $\pm 1.90$\,ps. For the 10\,ns frame interval, the largest TDC code is 155 and thus the average time uncertainty of an event is $155/2\times 1.90$\,ps\,$\approx 150$\,ps.
 Together with the detector timing jitter of 265\,ps FWHM, this leads to the measured spread of a time difference of $(2\,(265\,\text{ps})^2 + 2\,(150\,\text{ps})^2)^{1/2} \approx 430\,$ps FWHM. 
  
  The linear, triangular background originates from independent, i.e.~uncorrelated, sources which are dark counts and photons from different pairs. Their detection time is uniformly distributed in the measurement window. The distribution of the time difference between two of these uncorrelated events is therefore given by the convolution of two uniform distributions, leading to the triangular shape. By linear fitting and extrapolation, the accidental events can be removed from the signal of real coincidences (Fig.~\ref{fig:TimeDiffHisto}, solid line). The following results are obtained with a coincidence window $\Delta t = [-4,4]$ of 9~TDC units width and removed accidentals.

\begin{figure}[tp]
  \centering
	\includegraphics[scale=.8, trim=0.5cm 0 0 0]{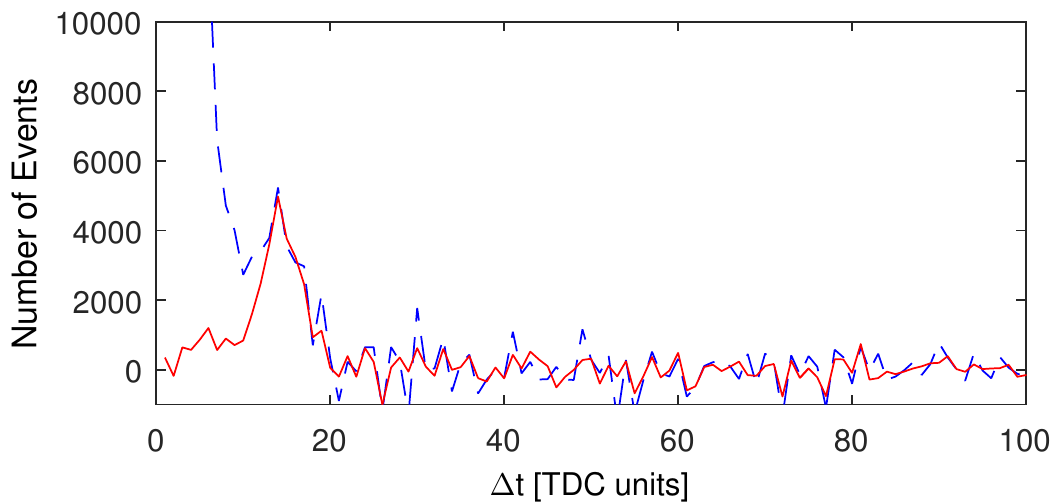}
    \caption{Histogram of the time difference $\Delta t$ between all events within each frame. One beam incident on the sensor is delayed by 300\,mm and photon pair detections are expected at $\Delta t = 15$\,TDC units. Coincidences between all pixels are considered in the dashed line where the crosstalk events at $\Delta t = 0$ TDC units rise to $4\times10^4$ events. In order to suppress these, only coincidences between the left half with the right half of the sensor array are taken into account in the solid line. Accidentals are removed in both graphs as shown in Fig.~\ref{fig:TimeDiffHisto} and 54\,M frames are evaluated.}
    \label{fig:TimeDiffHisto_delayed}
\end{figure}

The accidental corrected signal in Fig.~\ref{fig:TimeDiffHisto} contains not only photon pair coincidences but crosstalk events between neighbouring pixels as well. To experimentally confine the crosstalk and to  demonstrate temporal resolution, we increase the optical path delay between the right and the left beam incident on the sensor from 20\,mm to 300\,mm. The photon pair detection is therefore expected at $\Delta t = 15$\,TDC units. The effective temporal resolution limited by jitter and TDC variations together with still a high amount of crosstalk events at this $\Delta t$ leads to a masked coincidence signal (Fig.~\ref{fig:TimeDiffHisto_delayed}, blue line). Suppressing crosstalk by only considering events of pixels in the left half with those on the right half of the sensor (Fig.~\ref{fig:TimeDiffHisto_delayed}, red), a peak at $\Delta t=13$ TDC units with 5 units FWHM is visible. The path delay of 300\,mm corresponding to 1\,ns delay would lead to a TDC code difference of 15 units. A systematic difference of the TDC unit between pixels one the left half of the sensor with pixels on the right half explains this discrepancy.

\subsection{Spatial correlations}
\begin{figure}[t]
\centering
 \subfigure[]{
   \includegraphics[height=3.5cm]{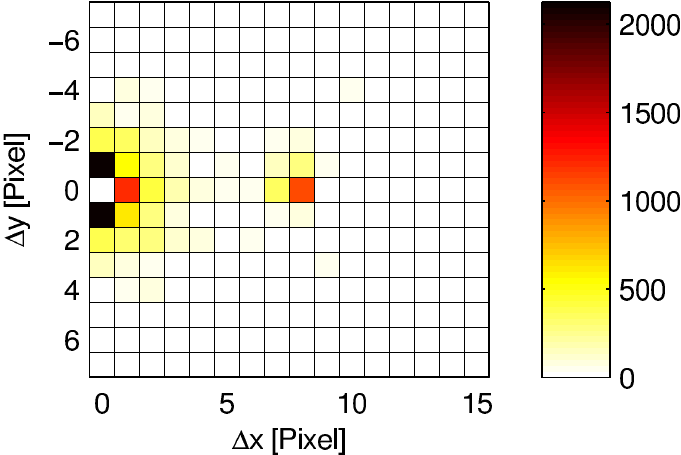}\label{fig:corrz0expt}}\hspace{1cm}
   \subfigure[]{
       \includegraphics[height=3.5cm]{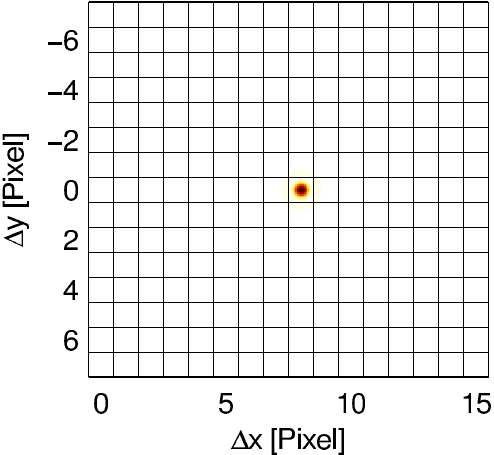}\label{fig:corrz0theory}}
   \caption{Second-order correlation function $G^{(2)}(\pmb{\Delta\varrho},0)$ of coincidence events in measurement (a) and theory (b). The NLC temperature is 25$^\circ$C, a coincidence window of 9\,TDC units is used and accidentals are removed.}
\end{figure}

We now spatially resolve the coincidence events in relative distances between two pixels using the difference coordinates $(\Delta x,\Delta y)$ in units of pixels (Fig.~\ref{fig:corrz0expt}). The experimental configuration is such that $z=0$\,mm, i.e.~the object plane coincides with the middle of the NLC, and the phase matching temperature is kept fix at $T=25^\circ$C. For $\Delta x = 8$\ pixels, which corresponds to the horizontal distance between the two beams, a narrow pixel correlation is observed. By means of the aforementioned values for $z$ and $T$, the second-order correlation function of Eq.~(\ref{eq:g2}) consistently shows a photon pair correlation width smaller than one pixel (Fig.~\ref{fig:corrz0theory}). A total of 2,372 events are registered in this region. This comes close to the theoretical value of 3,770 events 
which would be expected at the above measured PDE, the beam splitter ratio and the given photon flux. The discrepancy can be explained by losses in the optical setup which results in single photons without its partner.

The region centered at $\Delta x = 0$\,pixels comprises the coincidence events within each beam individually and includes a total of 11,176 events. The photon pair correlation width of less than one pixel measured at $\Delta x = 8$ pixels suggests, that photon pairs not separated by the beam splitter will be incident on the same pixel and therefore rarely lead to coincidence events between adjacent pixels. Hence, the central region is expected to show considerably less events, especially at a separation of $\Delta x \geq 2$. Therefore, most of the coincidence events have to be attributed to crosstalk between neighbouring pixels. A measurement with uncorrelated, classical light of similar power showed a comparable amount of events in this region (7,880 counts) and supports this conclusion.

\begin{figure}[t]
\centering
\subfigure[]{
   \includegraphics[height=3.5cm]{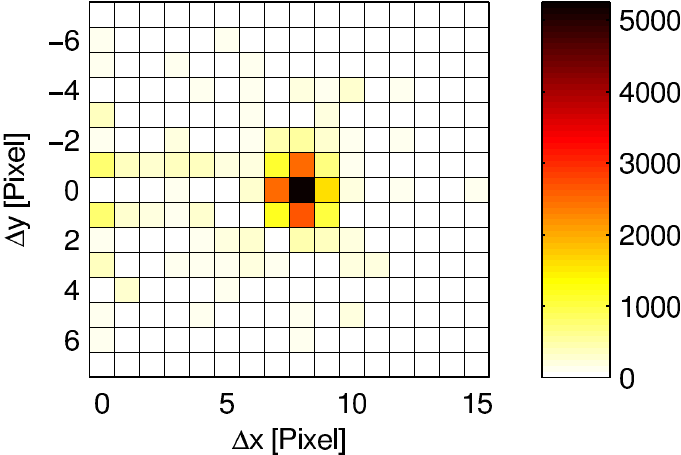}\label{fig:DeltaPos_taushift_exp}}\hspace{1cm}
 \subfigure[]{
	\includegraphics[height=3.5cm]{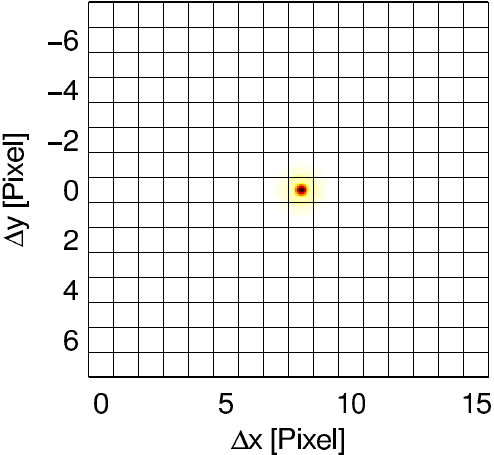}\label{fig:DeltaPos_taushift_the}}
   \caption{Second-order correlation function $G^{(2)}(\pmb{\Delta\varrho}, 0)$ of events with $\Delta t = 13\pm4$~TDC units in measurement (a) and theory (b). One beam is delayed by 300\,mm. Due to temporal separation  of the coincidence signal and crosstalk, the latter expected around $\Delta x = 0$ is suppressed (Fig.~\ref{fig:corrz0expt}). The NLC temperature is $25^\circ$C, 
    accidentals are removed and 54\,M frames are evaluated.  }
   \label{fig:DeltaPos_taushift}
\end{figure}

In order to suppress crosstalk events, the same procedure as in Section~\ref{AccRemoval} is applied. A path delay of 300\,mm between the left and the right beam temporally separates crosstalk and photon pair detection events. The second-order correlation function in Fig.~\ref{fig:DeltaPos_taushift_exp} shows strongly suppressed events around $\Delta x = 0$ in comparison to the measurement in Fig.~\ref{fig:corrz0expt}. The path delay leads to a defocussing and thereby enlarges the correlation function which results in a slightly broader correlation peak.

\begin{figure}[p]
    \centering
    \begin{subfigure}{}
        \includegraphics[scale=.7]{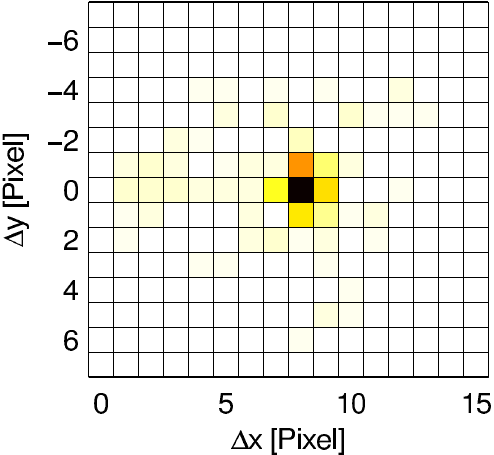}
    \end{subfigure}
    \hskip1em
    \begin{subfigure}{}
        \includegraphics[scale=.7]{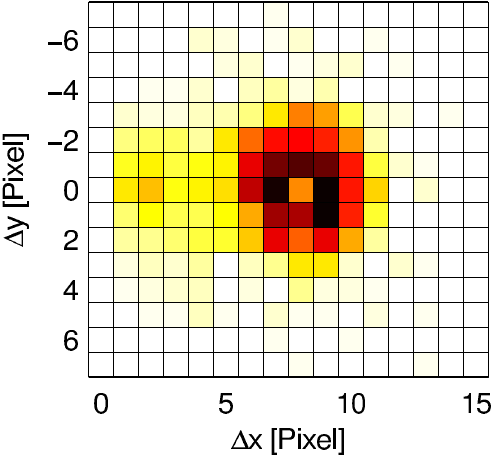}
    \end{subfigure}
    \hskip1em
     \begin{subfigure}{}
        \includegraphics[scale=.7]{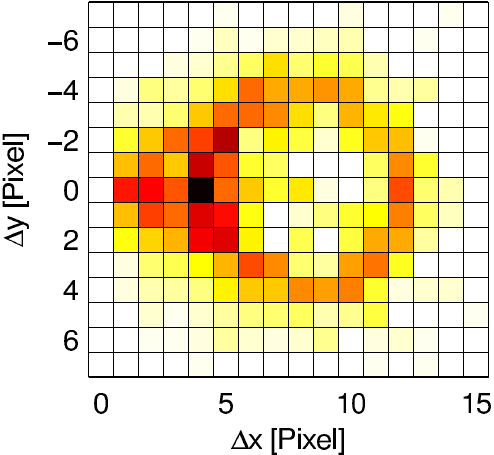}
    \end{subfigure}
    \vskip1em
    \begin{subfigure}{}
        \includegraphics[scale=.7]{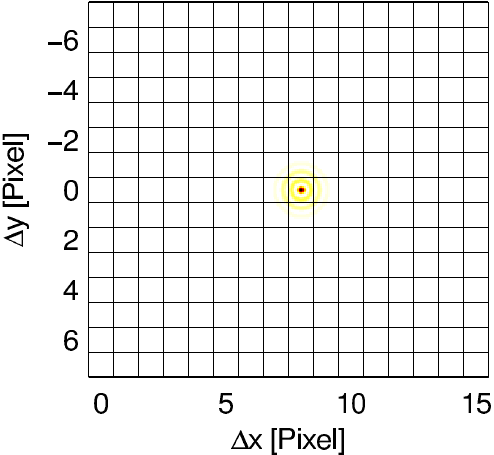}
    \end{subfigure}
    \hskip1em
    \begin{subfigure}{}
        \includegraphics[scale=.7]{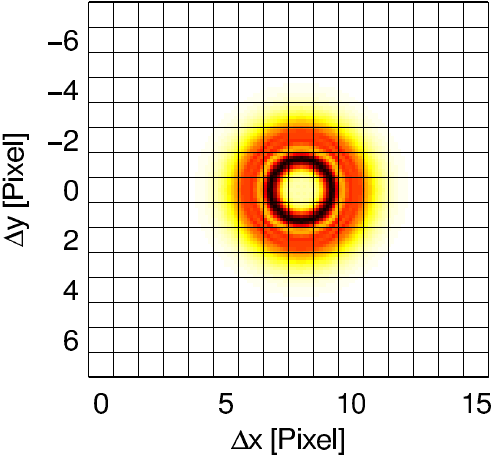}
    \end{subfigure}
    \hskip1em
    \begin{subfigure}{}
        \includegraphics[scale=.7]{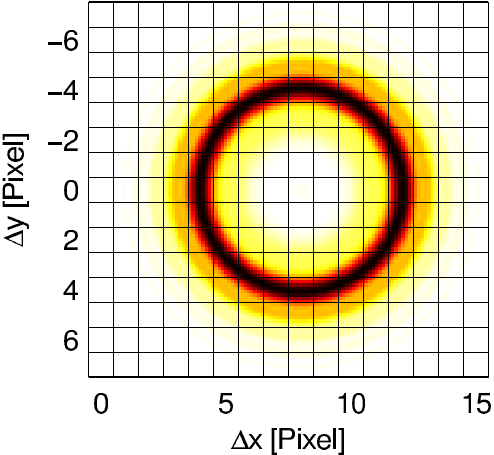}
    \end{subfigure}          
     \caption{\label{fig:G2Measurement_var_z}Measurements (upper row) and theory (lower row) of the  second-order correlation function $G^{(2)}(\pmb{\Delta\varrho},z)$ for different crystal positions $z = 0, 5, 10$\,mm and a fixed crystal temperatures $T=23^\circ C$. Pixel crosstalk is present at small $\Delta x$ and superimposes the light's coincidence signal.
      Every measurement consists of 54\,M frames. A coincidence window of 9~TDC units is used and accidentals are removed.}
\end{figure}

\begin{figure}[p]
    \centering
    \begin{subfigure}{}
        \includegraphics[scale=.7]{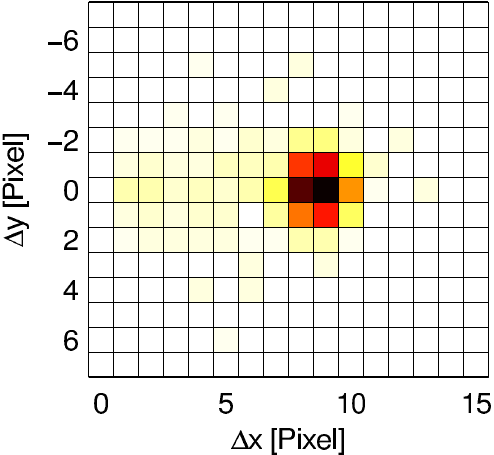}
    \end{subfigure}
    \hskip1em
    \begin{subfigure}{}
        \includegraphics[scale=.7]{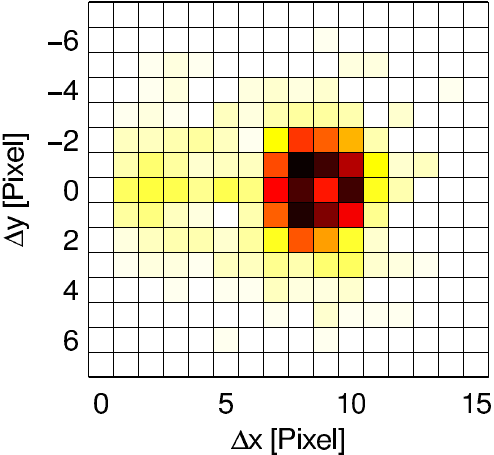}
    \end{subfigure}
    \hskip1em
     \begin{subfigure}{}
        \includegraphics[scale=.7]{G2_z5_T23_eps2eps-eps-converted-to.pdf}
    \end{subfigure}
    \vskip1em
    \begin{subfigure}{}
        \includegraphics[scale=.7]{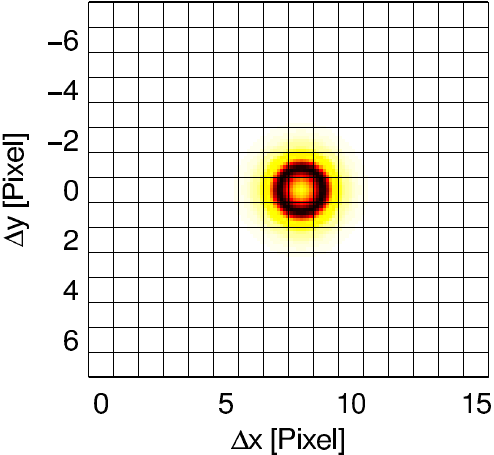}
    \end{subfigure}
    \hskip1em
    \begin{subfigure}{}
        \includegraphics[scale=.7]{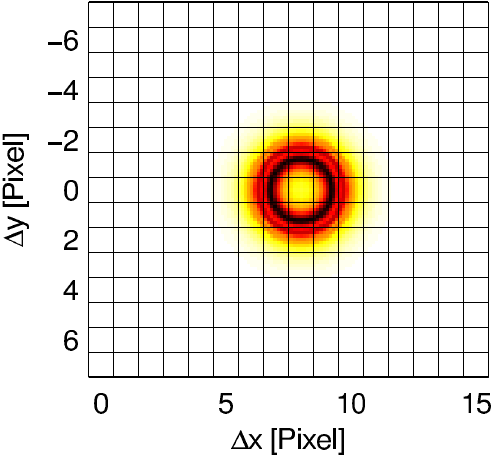}
    \end{subfigure}
    \hskip1em
    \begin{subfigure}{}
        \includegraphics[scale=.7]{G2_z5_T23_theory_eps2eps-eps-converted-to.pdf}
    \end{subfigure}          
     \caption{\label{fig:G2Measurement_var_T}Measurements (upper row) and theory (lower row) of the second-order correlation function $G^{(2)}(\pmb{\Delta\varrho},z)$ for fixed crystal position $z = 5$\,mm and different crystal temperatures $T = 25, 24, 23^\circ$C. Pixel crosstalk is present at small $\Delta x$ and superimposes the light's coincidence signal.
      Every measurement consists of 54\,M frames. A coincidence window of 9~TDC units is used and accidentals are removed. Note that in the first two measurements, the beam distance is slighty higher due to realignment.}
\end{figure}

The second-order correlation function shown in Fig.~\ref{fig:corrz0theory} has a width of approximately 0.3\,mm at FWHM which is not resolvable by a pixel of 0.6\,mm size. However, according to Eq.~(\ref{eq:g2}), the correlation function starts to broaden while moving the central plane of the NLC out of focus using the $z$ degree of freedom shown in Fig.~\ref{fig:expsetup}. In addition, lowering the crystal temperature $T$ allows to modify the SPDC phase matching from collinear to non-collinear emission of photon pairs which has a similar effect on the correlation function as changing the crystal's $z$-position.
By using 54\,M frames for better statistics (corresponding to 540\,ms effective measurement time acquired in 165\,s), 
Fig.~\ref{fig:G2Measurement_var_z} and Fig.~\ref{fig:G2Measurement_var_T} show experimental results in comparison with Eq.~(\ref{eq:g2}) for different settings of $z$ and $T$. As above, crosstalk is minimized by considering only coincident events of the left half with the right half of the sensor. In order to measure the correlation function in a plane, no path difference is introduced between the left and the right beam to temporally separate crosstalk as before. Because of that, crosstalk of the same magnitude is present in all measurements in the region of small $\Delta x$. Its relative strength increases the weaker the signal density gets from the first to the third column.

\clearpage

\section{Conclusions}

In this work, we demonstrate a high signal-to-noise ratio measurement of the spatial second-order correlation function of a high flux SPDC light source by means of a monolithic, fully digital and high temporal resolution SPAD pixel-array. The data acquisition time of the here presented measurements is below 3 minutes. Therefore, despite the low measurement duty cycle of 0.33\%, the simultaneous measurement on all pixels outweighs conventional scanning experiments in terms of mechanical complexity and in measurement time. This is even more the case for multi-photon experiments where higher  photon numbers are involved. With the given 128 pixels and second-order correlation measurements, a time reduction factor of $128^2\times0.33\%=54$ is realized compared to a single pixel scanning experiment with optimal duty cycle.
 
At a PDE of 0.57\% at 810\,nm and a total DCR of 7.9\,Mcps over all pixels, a high rate of accidental coincidence events originating from single photons and dark count events are expected. The photon arrival time resolution of 265\,ps allows to realize a small coincidence window which keeps accidentals at a minimum.
The remaining accidentals can be estimated very accurately and removed in a post-processing step.

The presented measurement of the second-order correlation function is currently only possible by splitting the photon pair and imaging it onto two distant parts of the sensor. Otherwise, pixel crosstalk would have superimposed and, to a large extent, masked the signal.
Since the digital signal handling prevents electrical crosstalk events, the observed crosstalk, over distances of more than 100\,$\mu$m, has to be mainly of photonic origin. Light emitted in SPAD avalanche events is reflected back by the glass surface of the chip and thereby leads to secondary detection events. Preliminary measurements with a sensor of the same type having no glass surface on top support this hypothesis. 

The sensor is currently not optimized for quantum imaging experiments, neither for the used wavelength and, compared to the sensors presented in \cite{burri14,krstajic15,lussana13,jahromi16,dutton16}, shows a considerably high DCR. Nevertheless, it showed the potential of a planned next generation sensor  where  a higher pixel density, a higher PDE and a better measurement duty cycle is desirable in order to further reduce the measurement time. The latter can be achieved by a higher frame rate or, to avoid unfeasible high data rates, by a frame readout triggered by multi-pixel events. To avoid the splitting into individual photon beams in future experiments, particular efforts will be put onto the reduction of crosstalk. This will pave the way for fast, higher photon number imaging experiments and applications.

\section*{Funding}
Swiss National Science Foundation (SNSF) (PP00P2\_159259); European Union's Horizon 2020 research and innovation programme (686731).



\end{document}